\documentclass[pre,twocolumn,superscriptaddress]{revtex4}

\usepackage{amsfonts,amsmath,amssymb,amsthm}
\usepackage{epic,eepic,graphicx,mathpple}
\usepackage{epstopdf}

\usepackage{amsmath, amsthm, amssymb}
\usepackage{graphicx}
\usepackage{caption}

\newtheorem{theorem}{Theorem}

\theoremstyle{definition}

\theoremstyle{remark}

\newcommand{\eeee}{\textrm{ee}}
\newcommand{\eeii}{\textrm{ei}}
\newcommand{\iiii}{\textrm{ii}}
\newcommand{\iiee}{\textrm{ie}}
\newcommand{\ee}{\textrm{e}}
\newcommand{\ii}{\textrm{i}}
\newcommand{\iin}{\textrm{in}}
\newcommand{\oout}{\textrm{out}}

\begin{document}

\keywords{}

\title{Highly connected neurons spike less frequently in balanced networks}

\author{Ryan Pyle}
\affiliation{Department of Applied and Computational Mathematics and Statistics, University of Notre Dame, Notre Dame IN 46556, USA}
\author{Robert Rosenbaum}
\affiliation{Department of Applied and Computational Mathematics and Statistics, University of Notre Dame, Notre Dame IN 46556, USA}
\affiliation{Interdisciplinary Center for Network Science and Applications, University of Notre Dame, Notre Dame IN 46556, USA}

\begin{abstract}
Many biological neuronal networks exhibit highly variable spiking activity. Balanced networks offer a parsimonious model of this variability.  In balanced networks, strong excitatory synaptic inputs are canceled by strong inhibitory inputs on average and spiking activity is driven by transient breaks in this balance.  Most previous studies of balanced networks assume a homogeneous or distance-dependent connectivity structure, but connectivity in biological cortical networks is more intricate.  We use a heterogeneous mean-field theory of balanced networks to show that heterogeneous in-degrees can break balance, but balance can be restored by heterogeneous out-degrees that are correlated with in-degrees.  In all examples considered, we find that highly connected neurons spike less frequently, consistent with recent experimental observations.
\end{abstract}

\maketitle

\section{Introduction}

Neuronal networks often exhibit highly irregular and asynchronous activity~\cite{Softky1993,Shadlen1994,Faisal2008} as well as a balance between positive (excitatory) and negative (inhibitory) interactions~\cite{Shu:2003ht,wehr:2003,Marino2005,Haider:2006gs,Okun:2008p752,Dorrn:2010hu,Sun:2010ih,Zhou2014,Petersen2014}.  Balanced network models offer a parsimonious model of this activity.  In balanced networks, chaotic or chaos-like dynamics produce irregular spiking activity through transient fluctuations in the balance of strong excitatory and inhibitory currents~\cite{vanVreeswijk:1996us,vanVreeswijk:1998uz,Renart:2004tn,Monteforte:2012hh,Lajoie2013}.  Most studies of balanced networks assume a homogeneous network architecture where connection probability depends only on cell polarity. This was recently extended to networks with distant-dependent connection probabilities~\cite{lim:2013,Rosenbaum2014}, but biological networks exhibit more diverse architectures~\cite{song2005highly,Sadovsky2013,Gururangan2014,LandauSfN}.  

In this article, we use heterogeneous mean-field theory to show that architectures with heterogeneous in-degree distributions and homogeneous out-degree distributions break the classical balanced state, consistent with a parallel study~\cite{LandauSfN}.  We next show that balance can be restored if out-degrees are also heterogeneous and correlated with in-degrees.  In each of the example architectures we consider, neurons with higher in-degrees have lower firing rates, consistent with recent experimental results showing a negative correlation between firing rate and local functional coupling strength in cortex~\cite{Okun2015}.


\section{Model description.} We consider a network of $N$ model neurons.  The membrane potential of neuron $j$ obeys  integrate-and-fire dynamics
$$
\frac{dV_j}{dt}=f(V_j)+I_j(t)
$$
with the added condition that each time $V_j(t)$ exceeds a threshold at $V_\textrm{th}$, the neuron spikes and the membrane potential is held for a refractory period $\tau_\textrm{ref}$ then reset to a fixed value $V_\textrm{re}$.  All simulations use the exponential integrate-and-fire (EIF) model~\footnote{EIF model defined by defined by $\tau_m f(V)=-(V-E_L)+\Delta_T \textrm{exp}[(V-V_T)/\Delta_T]$ with parameters $\tau_m=15\textrm{ms}$, $\Delta_T=2$~mV, $V_T=-55$~mV, $V_\textrm{th}=-50$~mV, $V_\textrm{re}=-75$~mV and $\tau_\textrm{ref}=0.5$~ms
}.  Synaptic input currents are defined by
\begin{equation}\label{E:input}
I_j(t)= \sum_{k=1}^{N}\frac{J_{jk}}{\sqrt N}\sum_n \alpha_k(t-t_{k,n})+\sqrt N F_j
\end{equation}
where, $t_{k,n}$ is the $n$th spike time of neuron $k=1,\ldots,N$.  Postsynaptic current waveforms, $\alpha_k(t)$, are assumed to satisfy $\alpha_k(t)=0$ for $t<0$ and, without loss of generality, $\int \alpha_k(t)dt=1$~\footnote{For all simulations, $\alpha_k(t)=(e^{-t/\tau_{d}}-e^{-t/\tau_{r}})/(\tau_{d}-\tau_{r})$ for $t>0$ with timescales $\tau_d=0.1$ and $\tau_r=6$ for excitatory presynaptic neurons and $\tau_d=0.1$ and $\tau_r=4$ for inhibitory neurons.
}.  
The term $F_j$ models feedforward input to the neuron from outside the network. 
Network structure is determined by the $N\times N$ matrix of connection strengths, $J$.  

We are interested in the statistics of network activity as $N$ grows large.   The $\sqrt N$ scaling of feedforward input and $1/\sqrt N$ scaling of synaptic weights are defining features of the balanced network formalism that permit chaotic, irregular spiking activity at large $N$~\cite{vanVreeswijk:1996us,vanVreeswijk:1998uz,Renart:2004tn,Monteforte:2012hh,Lajoie2013}.  





\section{Results}
\subsection{Heterogeneous mean-field theory of balanced networks.}
We first extend the mean-field theory of firing rates in balanced networks~\cite{vanVreeswijk:1996us,vanVreeswijk:1998uz,Renart:2004tn,Rosenbaum2014} to account for heterogeneous structure.   First partition the network of $N$ neurons into $K$ populations, where population $m$ contains $N_m$ neurons with $q_{m}=N_{m}/N\sim\mathcal O(1)$ for $m=1,\ldots, K$. 

Now define the average input to neurons in population $m$,
$$
\overline I_m=\textrm{avg}_{j\in G(m)}\left[I_j(t)\right]
$$
where $j\in G(m)$ indicates that the average is taken over all neurons in population $m$, and also over time.  Define $\overline F_m$ similarly and define $r_m$ to be the average spiking rate of neurons in population $m$.  Averaging Eq.~\eqref{E:input} over each population and over time gives the mean-field mapping 
\begin{equation}\label{E:WrF}
\vec I=\sqrt N\left(W\vec r+\vec F\right)
\end{equation}
where $\vec I=[\overline I_1\cdots \overline I_K]$ is the vector of mean inputs and similarly for $\vec r$ and $\vec F$.  The $K\times K$ mean-field connectivity matrix is defined by
$$
W=\left[q_n\overline J_{mn}\right]_{m,n=1}^{K}
$$
where 
$$
\overline J_{mn}=\frac{1}{N_m N_n}\sum_{\substack{j\in G(m),\,k\in G(n)}}J_{jk}
$$ 
is the average connection strength from neurons in population $n$ to neurons in population $m$, which is assumed to be $\mathcal O(1)$.

In the balanced state, $\vec r,\vec I\sim\mathcal O(1)$ as $N$ increases. 
From Eq.~\eqref{E:WrF}, however, this can only be achieved under a cancellation between positive and negative (excitatory and inhibitory) input sources in such a way that $W\vec r +\vec F\sim\mathcal O(1/\sqrt N)$.  This cancellation  defines the balanced network state.  As $N\to\infty$ firing rates are given by the solution to the balance equation
\begin{equation}\label{E:WrF0}
W\vec r+\vec F=0.
\end{equation} 
Thus, the existence of a balance state requires that Eq.~\eqref{E:WrF0} has a solution, $\vec r$, with positive components, $r_m>0$. 
When $W$ is invertible, this solution can be written as $\lim_{N\to\infty}\vec r=-W^{-1}\vec F$.
There are numerous ways to partition a network.   Thus, the solvability of Eq.~\eqref{E:WrF0} for a specific partition is a necessary, but not sufficient condition for the existence of a balanced state.  

The stability of the balanced state can be approximated by considering the dynamical mean-field equation~\cite{dayan2001theoretical,Rosenbaum2014} 
\begin{equation*}
\tau \vec r\,'=-\vec r+f\left(\sqrt N [W\vec r+\vec F]\right)
\end{equation*} 
where $f(\cdot)$ is a non-decreasing firing rate function applied element-wise and $\tau$ is a constant.  Assuming the gain, $f'(\vec I)$, at the balanced fixed point is $\mathcal O(1)$, the fixed point is stable as $N\to\infty$ whenever all eigenvalues of $W$ have negative real part~\cite{vanVreeswijk:1998uz}.




\subsection{A review of homogeneous balanced networks.} 
For the purpose of comparison, we first review networks with homogeneous connection probabilities that depend only on cell polarity (excitatory or inhibitory) as in~\cite{vanVreeswijk:1996us,vanVreeswijk:1998uz}.  For this model, $N_\ee=q_\ee N$ of the neurons are excitatory and $N_\ii=q_\ii N$ are inhibitory, where $q_\ee,q_\ii\sim\mathcal O(1)$.  All excitatory neurons receive the same feedforward input, $F_j=F_\ee>0$, and all inhibitory neurons receive $F_j=F_\ii>0$.  The synaptic connection strength, $J_{jk}$, from neuron $k$ in population $y=\ee,\ii$ to neuron $j$ in population $x=\ee,\ii$ are randomly assigned according to
$$
J_{jk}=\begin{cases}j_{xy} & \textrm{ with prob. }p_{xy}\\ 0 & \textrm{ otherwise}\end{cases}.
$$
Here, $p_{xy}$ represents the connection probability from population $y=\ee,\ii$ to population $x=\ee,\ii$ and $j_{xy}$ represents the strength of each the connection. Note that $j_\eeee,j_\iiee>0$ and $j_\eeii,j_\iiii<0$.

 \begin{figure}
{\centering{
\includegraphics[width=3.35in]{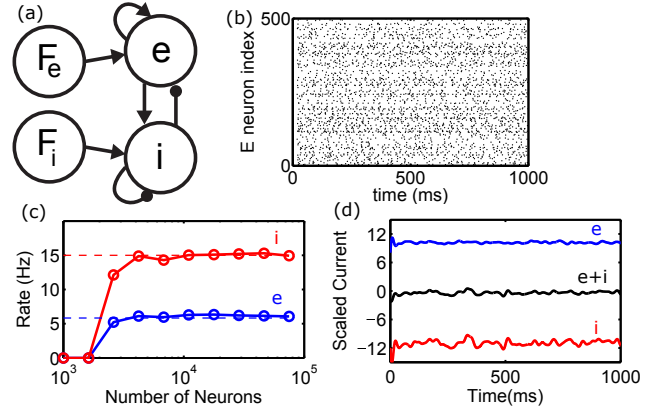}
}}
\caption{{\bf A homogeneous balanced network.} {\bf (a)} Network schematic.  A population of $N_\ee$ excitatory and $N_\ii$ inhibitory neurons ($\ee$ and $\ii$) are randomly connected and also receive feedforward input ($F_\ee$ and $F_\ii$).  {\bf (b)} Raster of plot of 500 randomly sampled excitatory neurons from a simulation of a balanced network with $N_\ee=4\times 10^4$ and $N_\ii=10^4$. {\bf (c)} Firing rates from simulations (solid curves) approach the values predicted by solving Eq.~\eqref{E:WrF0} (dashed lines) as network size, $N=N_\ee+N_\ii$, grows.  {\bf d)} Synaptic input to one representative excitatory neuron shows that strong excitatory currents (blue) balance with strong inhibitory currents (red) to yield a moderate total synaptic current (black).   Synaptic currents were convolved with a Guassian shaped filter ($\sigma=8$~ms) and normalized by the neuron's rheobase.
}
\label{F:Fig1}
\end{figure}

Dividing the network into excitatory and inhibitory populations and applying the mean-field theory outlined above gives the mean feedforward input, $\vec F=[F_\ee\; F_\ii]^T$.  Similarly, the mean-field connectivity matrix is given by 
\begin{equation}\label{E:Wh}
W_h=\left[\begin{array}{cc}w_\eeee & w_\eeii \\ w_\iiee & w_\iiii\end{array}\right]
\end{equation}
where 
$
w_{xy}=q_yp_{xy}j_{xy}
$ 
and the subscript $h$, for homogeneous, is used to distinguish this matrix from the ones we will consider below.  For this network, the balance equation \eqref{E:WrF0} has a stable, positive solution whenever~\cite{vanVreeswijk:1996us,vanVreeswijk:1998uz,Renart:2004tn,Rosenbaum2014}
\begin{equation}\label{E:HBalanceCond}
\frac{F_\ee}{F_\ii}>\frac{w_\eeii}{w_\iiii}> \frac{w_\eeee}{w_\iiee}.
\end{equation}
Computer simulations~\footnote{Parameters for all simulations: $j_\eeee=112.5$, $j_\eeii=-300$, $j_\iiee=225$, $j_\iiii=-450$, $F_\ee=0.0187$, $F_\ii=0.015$, $q_\ee=0.8$, $q_\ii=0.2$ and $p_{xy}=0.05$ for $x,y\in\{\ee,\ii\}$.} confirm the predicted firing rates and demonstrate the asynchronous, irregular spiking characteristic of the balanced state (Fig.~\ref{F:Fig1}).  We next show that re-wiring this network to produce heterogeneous in-degrees can break balance.


%



\subsection{Heterogeneous in-degrees can break balance.}

\begin{figure}
{\centering{
\includegraphics[width=3.35in]{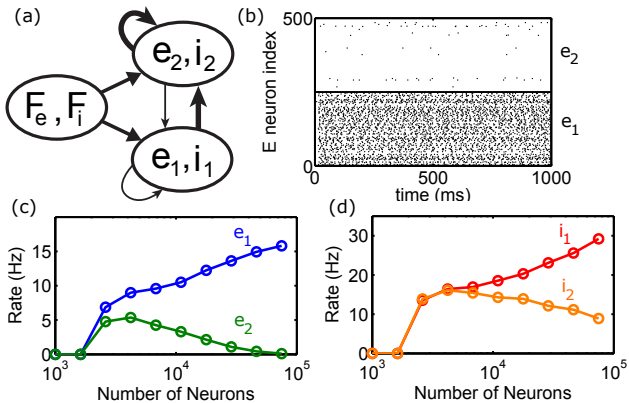}
}}
\caption{{\bf Heterogeneous in-degrees can break balance.} {\bf (a)} Network diagram.  
Same as the network from Fig.~\ref{F:Fig1} except the excitatory and inhibitory populations were each divided into two populations.  Neurons in populations $\ee_2$ and $\ii_2$ have larger in-degrees than those in populations $\ee_1$ and $\ii_1$.  {\bf (b)} Raster plot of 500 randomly selected excitatory neurons, half from $\ee_1$ and half from $\ee_2$, from a simulation with $N=5\times 10^4$ neurons.   {\bf (c,d)} Mean firing rate in each population as a function of network size ($N$).
}
\label{F:Fig2}
\end{figure}

As a first example of a heterogeneous network, we re-wired the homogeneous network above  to produce a bimodal distribution of in-degrees.  We first partitioned the excitatory population into two equal-sized sub-populations, $\ee_1$  and $\ee_2$.  We then did the same for the inhibitory population, giving a total of $K=4$ sub-populations which we enumerate as $\ee_1$, $\ii_1$, $\ee_2$ and $\ii_2$. 

A proportion $c_\iin=1/5$ of the incoming connections to postsynaptic neurons in populations $\ee_1$ and $\ii_1$ were randomly re-assigned to postsynaptic neurons in populations $\ee_2$ and $\ii_2$ respectively.  Thus, the average in-degrees of neurons in populations $\ee_2$ and $\ii_2$ were larger than those of neuron in populations $\ee_1$ and $\ii_1$ respectively (Fig.~\ref{F:Fig2}a).  The out-degrees and feedforward inputs were unchanged from Fig.~\ref{F:Fig1}.

In simulations of this network, the average firing rates of neurons in populations $\ee_1$ and $\ii_1$ were higher than the excitatory and inhibitory rates  in populations $\ee_2$ and $\ii_2$ (Fig.~\ref{F:Fig2}b-d).   Thus, perhaps surprisingly, a higher in-degree was associated with lower firing rates.   Increasing the network size while keeping connection probability fixed exaggerated this effect as firing rates in population $\ee_2$ approached zero (Fig.~\ref{F:Fig2}c,d).  

To understand this phenomenon intuitively, consider a simplified network diagram in which the populations with decreased in-degrees ($\ee_1$ and $\ii_1$) are grouped together (group 1) and those with increased in-degrees ($\ee_2$ and $\ii_2$) are also grouped together (group 2, Fig.~\ref{F:Fig2}a).  The increased in-degree of group 2 is then the equivalent of an increase in the mean strength of its self-connections and the mean strength of group-2-to-group-1 connections (indicated by thicker arrows in Fig.~\ref{F:Fig2}a).  

In the balanced state, strong inhibition cancels strong excitation, including excitatory feedforward input.  While both groups receive identical feedforward input, group 2 receives more recurrent input than group 1 regardless of the firing rates of each population.  Balance cannot be maintained in both groups because the same level of feedforward input received by each group cannot be simultaneously balanced by the two different levels of recurrent input they receive.  Group 2 receives an excess of inhibition because recurrent connections are net inhibitory in balanced networks~\cite{vanVreeswijk:1996us,vanVreeswijk:1998uz}, explaining why group 2 has lower firing rates than group 1.  

A more rigorous understanding is provided by applying the heterogeneous mean-field analysis described above.  The $4\times 1$ vector of mean feedforward inputs to populations $\ee_1$, $\ii_1$, $\ee_2$ and $\ii_2$ is given by $\vec F=[F_\ee\; F_\ii\; F_\ee\; F_\ii]^T$.  
The $4\times 4$ mean-field connectivity matrix is given in block form by
\begin{equation*}
W=\frac{1}{2}\left[\begin{array}{cc}  (1-c_\iin)W_h &  (1-c_\iin)W_h\\ (1+c_\iin)W_h & (1+c_\iin)W_h\end{array}\right]
\end{equation*}
where $W_h$ is the $2\times 2$ matrix from Eq.~\eqref{E:Wh}. 

Note that $W$ is singular and 
its range does not contain $\vec F$.  Thus, Eq.~\eqref{E:WrF0} does not admit a solution and this network re-wiring destroys balance. Only a non-generic choice of $\vec F$ within the range of $W$ could maintain balance.  For any other $\vec F$, firing rates in group 2 approach zero as $N\to\infty$ due to an excess of recurrent inhibition. Thus, re-wiring a homogeneous network to achieve heterogeneous out-degrees can destroy balance~\cite{LandauSfN}, causing highly connected sub-populations to cease spiking.   We next show that balance can be restored by heterogeneous out-degrees that are correlated with in-degrees.



%

\subsection{Balance can be restored by heterogeneous out-degrees.}

The re-wiring of the homogeneous network from Fig.~\ref{F:Fig1} considered above only altered in-degrees of neurons.  Starting from this rewiring, we now also change the out-degrees by rewiring the source of some edges.  Specifically, a proportion $c_\oout=4/5$ of the synaptic projections from presynaptic neurons in population $\ee_1$ to postsynaptic neurons in population $\ee_2$ are rewired to emanate from randomly selected presynaptic neurons in population $\ee_2$, \emph{i.e.} they now project from $\ee_2$ to $\ee_2$.  Similarly, a proportion $c_\oout=4/5$ of projections from neurons in $x_1$ to neurons in $y_2$ are rewired to form $x_2$-to-$y_2$ projections for all pairings of $x,y\in\{\ee,\ii\}$.

\begin{figure}
{\centering{
\includegraphics[width=3.35in]{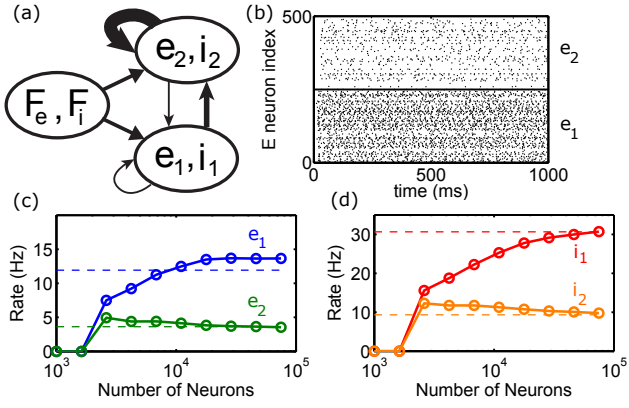}
}}
\caption{{\bf Balance can be restored by heterogeneous out-degrees.} Same as Fig.~\ref{F:Fig2}, except that the out-degrees of neurons in populations $\ee_2$ and $\ii_2$ were increased by rewiring a proportion $c_\oout=4/5$ of the outgoing projections from populations $\ee_1$ and $\ii_1$ to project from $\ee_2$ and $\ii_2$ instead.  Dashed lines shows the asymptotic firing rates predicted by Eq.~\eqref{E:WrF0}.
}
\label{F:Fig3}
\end{figure}

This rewiring  increases the average out-degree of neurons in populations $\ee_2$ and $\ii_2$ by a proportion $c_\oout$ and decreases the out-degrees of neurons in population $\ee_1$ and $\ii_1$ by the same proportion.  Since $\ee_2$ and $\ii_2$ also have larger in-degrees, this results in positively correlated in- and out-degrees (Fig.~\ref{F:Fig3}a).  

Simulating this network, we found that the average firing rates of neurons in populations $\ee_1$ and $\ii_1$ were larger than the rates  in populations $\ee_2$ and $\ii_2$ respectively  (Fig.~\ref{F:Fig3}b-d), but the difference was less drastic than the example with just heterogeneous in-degrees (compare to Fig.~\ref{F:Fig2}).   Increasing the network size while keeping connection probability fixed caused rates to approach non-zero limits (Fig.~\ref{F:Fig3}c,d).

We again analyze this network using the heterogeneous mean-field theory described above.  Using the same partition of the network into sub-populations $\ee_1$, $\ii_1$, $\ee_2$ and $\ii_2$, the mean-field feedforward input vector is again given by $\vec F=[F_\ee\; F_\ii\; F_\ee\; F_\ii]^T$ and 
the $4\times 4$ mean-field connectivity matrix is given in block form by
$$
W=\frac{1}{2}\left[\begin{array}{cc} (1-c_\iin)W_h &  (1-c_\iin)W_h\\ (1+c_\iin)(1-c_\oout)W_h & (1+c_\iin)(1+c_\oout)W_h\end{array}\right]
$$
where $W_h$ is from Eq.~\eqref{E:Wh}.  
It is easily shown that this network admits a stable, balanced state (\emph{i.e.}, Eq.~\eqref{E:WrF0} has positive solutions and $\det(W)\le 0$) if Eq.~\eqref{E:HBalanceCond} is satisfied and $c_\oout>c_\iin(2-c_\oout)$.  For the example considered here, this is satisfied and the balanced firing rates given by $\vec r=-W^{-1}\vec F$ agree with network simulations (Fig.~\ref{F:Fig3}c,d).  



This example shows that heterogeneous out-degrees can recover the balance lost by heterogeneous in-degrees.   Moreover, neurons with stronger incoming recurrent connections spike less frequently in this example.  The following theorem generalizes this finding.
\begin{theorem}
Consider any network with inputs defined by Eq.~\eqref{E:input} that realizes a stable balanced state.  If the network is broken into two groups such that the average feedforward input to each group is the same and positive, then the group with the larger average incoming recurrent connection strength (if there is one) has a smaller average firing rate.
\end{theorem}
The proof of this theorem is given in the Supplementary Materials.  Note that it is  possible for the highly connected group to have larger firing rates if each group receives different levels of feedforward input on average.  

\subsection{Firing rates in a scale free network}

So far we  considered networks with a finite number of groups where neurons within a group are statistically identical.  Many classes of networks cannot be easily captured with such a model.  We next investigate whether our finding that rates are lower for highly connected neurons is valid in a scale-free network with power-law degree distributions.


We assign to each neuron an in-degree, $u$, drawn independently from a generalized Pareto distribution with probability density function
$$
Q(u)= \begin{cases}\frac{1}{\sigma}\left(1+\frac{u-\mu}{\sigma}\xi\right)^{-(\xi^{-1}-1)} & u\ge \mu\\ 0 & u<\mu\end{cases}
$$
with shape parameter $\xi=0.25$, location parameter $\mu=5$ and scale parameter,
$$
\sigma=(\overline pN-\mu)(1-\xi)
$$
giving an average connection probability, $\overline p=0.05$.   In-degrees are then instantiated by drawing $\textrm{round}(u)$ excitatory and inhibitory presynaptic neurons randomly and uniformly from the network.  Thus, in-degrees obey a power-law distribution, but out-degrees are homogeneous.  Feedforward input strengths depend only on cell polarity, as above.

\begin{figure}
{\centering{
\includegraphics[width=3.35in]{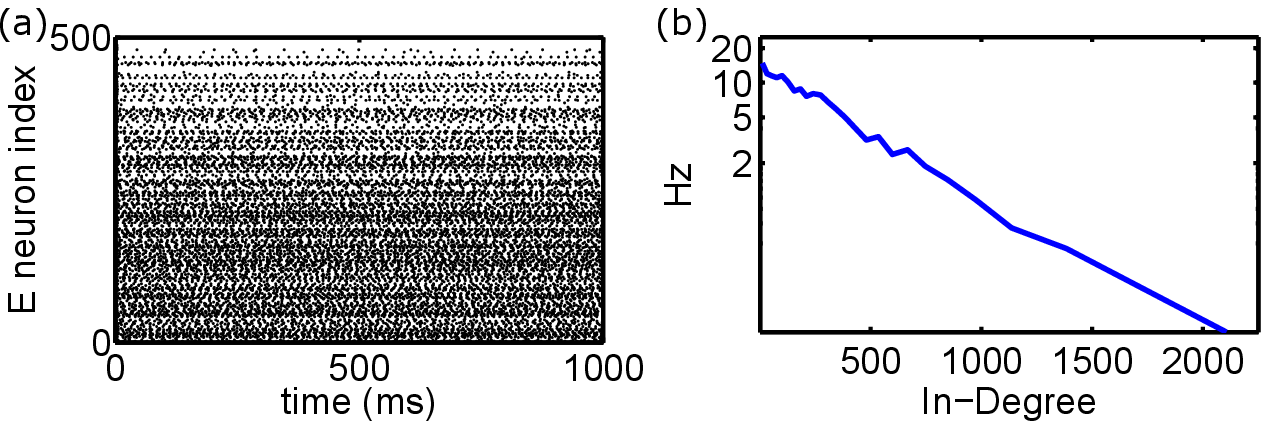}
}}
\caption{{\bf Dependence of firing rates on in-degree in a scale free network.} {\bf (a)} Raster plot and {\bf (b)} firing rates as a function of in-degree from a network of $5\times 10^4$ neurons with a power-law distribution of in-degrees.  For the raster plot, 500 excitatory neurons were sampled uniformly from the network and sorted so that in-degree increased with ``neuron index.''  
}
\label{F:Cts}
\end{figure}

Simulating this network confirms that firing rates are lower for neurons with higher in-degree (Fig.~\ref{F:Cts}), analogous to the networks considered above. 

The heterogeneous mean-field analysis outlined above can be applied by partitioning the network according to in-degree and neuron polarity.  In the limit of large $N$ and finer partitions, the matrix equation \eqref{E:WrF0} is approximated by a system of integral equations (compare to spatial networks in~\cite{Rosenbaum2014}),
\begin{equation}\label{E:Wint}
\begin{aligned}
&\int_\mu^\infty [w_\eeee(u,v)r_\ee(v)-w_\eeii(u,v)r_\ii(v)]dv+F_\ee =0\\
&\int_\mu^\infty [w_\iiee(u,v)r_\ee(v)-w_\iiii(u,v)r_\ii(v)]dv +F_\ii =0.
\end{aligned}
\end{equation}
Here, $r_x(v)$ is the average firing rate of neurons in population $x=\ee,\ii$ with in-degree $\textrm{round}(v)$.  The term
$$
w_{xy}(u,v)=Q(v)j_{xy}p(u,v)
$$ 
represents  mean-field connectivity from neurons in population $y=\ee,\ii$ with in-degree $v$ to neurons in population $x=\ee,\ii$ with in-degree $u$ where $p(u,v)$ represents the probability and $j_{xy}$ the strength of such a connection.  For the example considered here, connection probability depends only on the in-degree of the post-synaptic neuron so that $p(u,v)= u/N$. Note that $u\sim\mathcal O(N)$ so that $p(u,v)\sim\mathcal O(1)$ on average.  Thus Eqs.~\eqref{E:Wint} can be written as
$$
\begin{aligned}
&\frac{u}{N}\left[ j_\eeee\overline r_\ee- j_\eeii\overline r_\ii\right]+F_\ee =0\\
&\frac{u}{N}\left[ j_\iiee \overline r_\ee- j_\iiii \overline r_\ii\right]+F_\ii =0
\end{aligned}
$$
where $\overline r_x=\int Q(v)r_x(v)dv$ is the average firing rate of neurons in population $x=\ee,\ii$.  For balance to be achieved, this equation must be satisfied simultaneously for all $u>\mu$, which is not possible.  We conclude that the network in Fig.~\ref{F:Cts} violates the classical balanced state, like the example in Fig.~\ref{F:Fig2}.

Restoring balance in this example would require building a family of networks indexed by $N$, where connection probability, $p(u,v)\sim\mathcal O(1)$, depends on pre- and post-synaptic in-degree, $u$ and $v$, in such a way that Eqs.~\eqref{E:Wint} are solvable with $r_x(v)\ge 0$.  
Specifying and generating such a network is non-trivial and outside the scope of this study, but warrants further study.

\section{Discussion}

We used heterogeneous mean-field theory to analyze structured balanced networks.  Similar to the theory of homogeneous and spatially-extended balanced networks, firing rates in the limit of large network size are determined by a linear equation involving only synaptic parameters~\cite{vanVreeswijk:1996us,vanVreeswijk:1998uz,Renart:2004tn,Rosenbaum2014}.  The solvability of this equation determines the existence of the balanced state in the thermodynamic limit. 

We found that heterogeneous in-degrees destroy balance, but balance can be recovered by heterogeneous out-degrees.  In all heterogeneous networks we considered, neurons with higher in-degrees had lower firing rates.  This could potentially explain the negative correlation between firing rate and local population coupling recently observed in cortical recordings~\cite{Okun2015}.

It is possible to create balanced networks where neurons with high in-degree have higher rates.  All such examples are captured by Eq.~\eqref{E:WrF0}.  However, we expect that higher in-degrees will typically promote lower firing rates because recurrent input is net-inhibitory in balanced networks~\cite{vanVreeswijk:1996us,vanVreeswijk:1998uz,Renart:2004tn}.


%

The imbalance created by heterogeneous in-degrees quenches firing rates in some neurons and increases rates in others.  This effect is only realized at sufficiently large $N$.  Biological networks are, of course, finite in size.  At sufficiently small $N$, rates can be positive even if Eq.~\eqref{E:WrF0} has no solution (as in Fig.~\ref{F:Fig2}).  Firing rates in such finite sized networks could potentially be approximated  numerically using a diffusion approximation that gives a system of non-linear fixed point equations~\cite{Amit:1997uj,Renart:2004tn}.  
Once $N$ is large enough to quench spiking in some sub-populations, the remaining sub-populations might form a balanced sub-network that approaches finite rates at large $N$.  However, specifying this sub-network requires knowledge of which sub-populations have their rates quenched first, which in turn requires the analysis of finite-sized networks.

Inhibitory-stabilized networks, of which balanced networks are a special case, often exhibit net-inhibitory recurrent input~\cite{Ozeki2009}. We therefore expect that highly connected neurons will tend to spike less frequently in this more general class of network models.

A parallel study also reached the conclusion that balance is broken by heterogeneous in-degrees, but recovered balance through a strong adaptation current, which acts like a self-inhibition~\cite{LandauSfN}.  This resolution requires that adaptation currents are $\mathcal O(\sqrt N)$ to cancel excess synaptic input.  This might be a reasonable assumption at the finite network sizes of biological networks.


Previous studies consider recurrent neuronal networks with various types of heterogeneous connectivity structures~\cite{zhao2011synchronization,beggs2003neuronal}, but not in the balanced state.  Future work will consider the application of our balanced mean-field theory to these alternative architectures.

\section{Acknowledgements}
This work was supported by NSF-DMS-1517828.

%


\end{document}